\documentclass[prd,reprint,showkeys]{revtex4}%
\usepackage{amssymb}
\usepackage{amsfonts}
\usepackage{amsmath}
\usepackage{graphicx}%
\providecommand{\U}[1]{\protect\rule{.1in}{.1in}}

\begin{document}

\title{Stationary Scalar Clouds Around Maximally Rotating Linear Dilaton
Black Holes}

\author{G. Tokgoz}
\email{gulnihal.tokgoz@emu.edu.tr}

\author{I. Sakalli}
\email{izzet.sakalli@emu.edu.tr}

\date{\today }
\begin{abstract}
We investigate the wave dynamics of a charged massive scalar field
propagating in a maximally rotating (extremal) linear dilaton black
hole geometry. We prove the existence of a discrete and infinite family
of resonances describing non-decaying stationary scalar configurations
(clouds) enclosing these rapidly rotating black holes. The results
obtained signal the potential stationary scalar field distributions
(dark matter) around the extremal linear dilaton black holes. In particular,
we analytically compute the effective heights of those clouds above
the center of the black hole. 
\end{abstract}


\maketitle

\affiliation{Physics Department , Eastern Mediterranean University, Famagusta,
Northern Cyprus, Mersin 10, Turkey}

\section{Introduction}

According to the ``no-hair conjecture\textquotedblright \ \cite{Wheeler,Book1,Hod1},
which is one of the milestones in understanding the subject of black
hole (BH) physics \cite{Book2,Book3}, BHs are fundamental objects
like the atoms in quantum mechanics, and they should be characterized
by only 3-parameters: mass, charge, and angular momentum. In fact,
the no-hair conjecture has an oversimplified physical picture: all
residual matter fields for a newly born BH would either be absorbed
by the BH or be radiated away to spatial infinity (this scenario excludes
the fields having conserved charges) \cite{Hod1,BekT2,HodSr,Her1,Her2}.
In accordance with the same line of thought, other no-hair theorems
indeed omitted static spin-$0$ fields \cite{sp01,sp02,sp03,sp04,sp05},
spin-$1$ fields \cite{sp11,sp12,sp13,sp14}, and spin-$\frac{1}{2}$
fields \cite{sph1,sph2} from the exterior of stationary BHs. On the
other hand, significant developments in theoretical physics has led
to other types of ``hairy\textquotedblright \ BH solutions. A notable
example is the colored BHs in Refs. \cite{CBH1,CBH2}. In addition
to mass, for full characterization, a colored BH needs an additional
integer number (independent of any conserved charge) which is assigned
to the nodes of the Yang-Mills field. Other hairy BHs include different
types of exterior fields that belong to the Einstein-Yang-Mills-Dilaton,
Einstein-Yang-Mills-Higgs, Einstein-Klein-Gordon, Einstein-Skyrme,
Einstein-Non-Abelian-Proca, Einstein-Gauss-Bonnet etc. theories (see
for example, \cite{HBH17,HBH172,HBH173,HBH1,HBH2,HBH3,HBH4,HBH5,HBH6,HBH7,HBH8,HBH9,HBH10,HBH11,HBH12,HBH13,HBH14,HBH15,HBH16,HBH18,HBH19,HBH20,Bernard}).

Interestingly, many no-hair theories \cite{BekT2,sp13,sp14,sph1,sph2,NH1,NH4,NH6,NH7,NH8}
do not cover the time-dependent field configurations surrounding the
BH. Astronomical BHs are not tiny and unstable, but very heavy, large,
and practically indestructible. Observations show that in the densely
populated center of most galaxies, including ours, there are monstrous
BHs \cite{GALAXY}, which are many hundreds of millions of times heavier
than the sun. As was shown in \cite{Barro}, the regular time-decaying
scalar field configurations surrounding a supermassive Schwarzschild
BH do not fade away in a short time (according to the dynamical chronograph
governed by the BH mass). Besides, ultra-light scalar fields are considered
as a possible candidate for the dark matter halo (see for instance,
\cite{Barro,DM1,DM2,DM3}).

Hod \cite{HodSr} extended the outcomes of \cite{Barro} to the Kerr
BH, which is well-suited for studying astrophysical wave dynamics.
Hod proved the presence of an infinite family of resonances (discrete)
describing non-decaying (stationary) scalar configurations surrounding
a maximally rotating Kerr BH. Thus, contrary to the finite lifetime
of the static regular scalar configurations mentioned in \cite{Barro},
the\textit{\ }stationary and regular scalar field configurations
(\textit{clouds})\ surrounding the realistic rotating BH\ (Kerr)
survive infinitely long \cite{HodSr}. To this end, Hod considered
the dynamics of massive Klein-Gordon equation in the Kerr geometry.
Moreover, the effective heights of those clouds above the center of
the BH were analytically computed. The obtained results support the
lower bound conjecture of N\'{u}\~{n}iez, et al. \cite{NH6}.

In line with the study by \cite{HodSr}, the purpose of the present
study is to explore the\textit{\ }stationary and regular scalar field
configurations surrounding maximally rotating linear dilaton BH (MRLDBH)
\cite{Clement}. These BHs have a non-asymptotically flat structure,
similar to our universe model: Friedmann-Lema\^{i}tre-Robertson-Walker
spacetime \cite{universe}. Rotating linear dilaton BHs are the solutions
to the Einstein-Maxwel-dilaton-axion theory \cite{Leygnac}. These
BHs include dilaton and axion fields, which are canditates for the
dark matter halo. Some current experiments are focused on the relationship
between the dark matter and the dilaton and axion fields \cite{exp1,exp2,exp3,exp4}.
Various studies have also focused on the rotating linear dilaton BHs
\cite{rbh0,ldbh1,ldbh2,ldbh3,ldbh4,ldbh5}. In particular, the problem
of area quantization (see \cite{area}\ for the insights of the famous
Bekenstein's area conjecture) from boxed quasinormal modes, which
are obtained from caged massless scalar clouds, has been recently
studied in \cite{ldbh4}. Our present study considers the charged
massive Klein-Gordon-Fock (KGF) equation \cite{Klein,Gordon,Fock}
in the MRLDBH geometry. Thus, we investigate wave dynamics in that
geometry and seek for the existence of possible resonances describing
the stationary charged and massive scalar field configurations surrounding
the MRLDBH.

This paper is organized as follows. In Sec. II, we review the rotating
linear dilaton BHs with their characteristic properties and study
the charged scalar field perturbation in this geometry. In Sec. III,
we explore the existence of a discrete family of resonances describing
stationary scalar configurations surrounding MRLDBH, and in sequel
we compute the effective heights of those scalar configurations above
the central MRLDBH. Finally, we provide conclusions in Sec. IV. We
use the natural units with $c=G=k_{B}=\hbar=1$.

\section{Rotating linear dilaton BH spacetime and separation of KGF equation}

In this section, we consider a charged massive scalar field coupled
to a rotating linear dilaton BH. For a comprehensive analytical study,
the rotating linear dilaton BH is assumed to be extremal, which requires
the equality of mass term ($M$) with the rotation term ($a$). Namely,
we focus on the case of a charged massive test scalar field in the
geometry of MRLDBH spacetime.

The action of the Einstein-Maxwell-dilaton-axion theory is given by
\cite{Leygnac} 
\begin{equation}
S=\frac{1}{16\pi}\int\sqrt{\left\vert g\right\vert }\left\{ R-\frac{1}{2}e^{-4\phi}\partial_{\mu}\aleph\partial^{\mu}\aleph-2\partial_{\mu}\phi\partial^{\mu}\phi-\aleph F_{\mu\nu}\widetilde{F}^{\mu\nu}-e^{-2\phi}F_{\mu\nu}F^{\mu\nu}\right\} d^{4}x,\label{1}
\end{equation}
\qquad{}

where $R$ is the Ricci scalar, $F_{\mu\nu}$ denotes the Maxwell
tensor (antisymmetric rank-2 tensor field), and $\widetilde{F}^{\mu\nu}$
represents the dual of $F_{\mu\nu}$. Besides, $\phi$ and $\aleph$
are the dilaton field and the axion field (pseudoscalar), respectively.
The metric solution to action (1) is designated with the rotating
linear dilaton BH \cite{Clement}, which is described in the Boyer-Lindquist
coordinates as follows:

\begin{equation}
ds^{2}=-f(r)dt^{2}+\frac{dr^{2}}{f(r)}+h(r)\left[d\theta^{2}+\sin^{2}\theta\left(d\varphi-\frac{a}{h(r)}dt\right)^{2}\right].\label{2}
\end{equation}
The metric functions are given by 
\begin{equation}
h(r)=rr_{0},\label{3}
\end{equation}
\begin{equation}
f(r)=\frac{Z}{h(r)},\label{4}
\end{equation}

where the constant parameter $r_{0}$ is directly proportional to
the background electric charge $Q$: $r_{0}=\sqrt{2}Q$. In Eq. (4),
$Z=\left(r-r_{2}\right)\left(r-r_{1}\right)$, where, $r_{1}$ and
$r_{2}$ are the two positive roots of the condition of $f(r)=0$.
In fact, $r_{1}$ and $r_{2}$ radii represent the inner and outer
horizons, respectively. The explicit forms of those radii are given
by

\begin{equation}
r_{1}=M-\sqrt{M^{2}-a^{2}},\label{5n}
\end{equation}

\begin{equation}
r_{2}=M+\sqrt{M^{2}-a^{2}}.\label{6n}
\end{equation}

In fact, $M$ is an integration constant in deriving the rotating
linear dilaton BH solution. It is twice the quasilocal mass $M_{QL}$
\cite{BROWN-YORK} of this non-asymptotically flat BH: $M=2M_{QL}$.
The rotation parameter $a$ is related with the angular momentum ($J$)
of the rotating linear dilaton BH via $a=\frac{\sqrt{2}J}{Q}$. Meanwhile,
it is obvious from Eqs. (5) and (6) that having a BH solution, one
should impose the condition of $M\geq a$. Thus, MRLDBH geometry corresponds
to $a=M$.

The dilaton and axion fields are governed by \cite{Clement}

\begin{equation}
e^{-2\phi}=\frac{h\left(r\right)}{r^{2}+a^{2}\cos^{2}\theta},\label{7}
\end{equation}

\begin{equation}
\aleph=-\frac{\sqrt{2}Qa\cos\theta}{r^{2}+a^{2}\cos^{2}\theta}.\label{8}
\end{equation}

Moreover, the electromagnetic 4-vector potential is given by 
\begin{equation}
A^{em}=\frac{1}{\sqrt{2}}\left(e^{-2\phi}dt+a\sin^{2}\theta d\varphi\right).\label{9}
\end{equation}

The Hawking temperature \cite{WALD,ROVELLI} of the rotating linear
dilaton BH can be obtained from the definition of surface gravity
$\kappa$, as follows: 
\begin{equation}
T_{H}=\frac{\kappa}{2\pi}=\frac{1}{4\pi}\left.\frac{df(r)}{dr}\right\vert _{r=r_{2}}=\frac{r_{2}-r_{1}}{4\sqrt{2}Q\pi r_{2}}.\label{10}
\end{equation}

As can be seen from Eq. (10), in the case of maximal rotation ($M=a \rightarrow r_{1}=0$: the extreme case), the rotating linear dilaton
BH emits radiation with a constant temperature (since $Q$ possesses
a fixed value), which is independent from the mass: $T_{H}=\left(4\sqrt{2}\pi Q\right)^{-1}$.
Such a radiation is nothing but the well-known isothermal process
in the subject of the thermodynamics. If $A_{H}$\ stands for the
surface area of the event horizon of the rotating linear dilaton BH,
then the entropy of this BH is given by 
\begin{equation}
S_{BH}=\frac{A_{H}}{4}=\sqrt{2}\pi Qr_{2}.\label{11}
\end{equation}

Angular velocity of the rotating linear dilaton BH is expressed as

\begin{equation}
\Omega_{H}=-\left.\frac{g_{tt}}{g_{t\varphi}}\right\vert _{r=r_{2}}=\frac{J}{Q^{2}r_{2}}.\label{12}
\end{equation}

Thus, the first law of thermodynamics of the rotating linear dilaton
BH is evincible through the following differential equation 
\begin{equation}
dM_{QL}=T_{H}dS_{BH}+\Omega_{H}dJ.\label{13}
\end{equation}
It is worth noting that Eq. (13) does not involve the electrostatic
potential $\Phi_{e}$ \cite{GenThrm} since $Q$ represents the fixed
background charge \cite{Clement}.

The dynamics of a charged massive scalar field $\Psi$ in the rotating
linear dilaton BH spacetime is governed by the KGF equation (see for
example, \cite{GOHAR}):

\begin{equation}
\left(\partial_{\mu}-iqA_{\mu}^{em}\right)\left(\sqrt{-g}g^{\mu\nu}\left(\partial_{\nu}-iqA_{\nu}^{em}\right)\Psi\right)-\sqrt{-g}\mu^{2}\Psi=0,\label{14}
\end{equation}

where $q$ and $\mu$ are the charge and mass of the scalar particle,
respectively. We assume the ansatz for $\Psi,$ as follows:

\begin{equation}
\Psi\equiv\Psi_{lm}\left(t,r,\theta,\varphi\right)=e^{im\varphi}S_{lm}(\theta)R_{lm}(r)e^{-i\omega t},\label{15}
\end{equation}

where $\omega$ is the conserved frequency of the mode, and $l$ and
$m$ are the spheroidal and azimuthal harmonic indices, respectively,
with $-l\leq m\leq l$. In Eq. (15), $R_{lm}$ and $S_{lm}$ are the
functions of radial and angular equations of the confluent Heun differential
equation with the separation constant $\lambda_{lm}$ \cite{Heun,Heun1,Iz2016}.

For the MRLDBH spacetime ($M=a$), the angular part of Eq. (14) obeys
the following differential equation of spheroidal harmonics $S_{lm}(\theta)$
\cite{ABRAM,TEU,TEU2}: 
\begin{equation}
\frac{1}{\sin\theta}\frac{\partial}{\partial\theta}\left(\sin\theta\frac{\partial S_{lm}}{\partial\theta}\right)+\left[\lambda_{lm}-\left(\frac{1}{2}\widehat{q}M\sin\theta\right)^{2}-\frac{m^{2}}{\sin^{2}\theta}\right]S_{lm}=0.\label{16}
\end{equation}

where $\widehat{q}=\sqrt{2}q.$ The above differential equation has
two poles at $\theta=0$ and $\theta=\pi$. For a physical solution,
$S_{lm}$ functions are required to be regular at those poles. This
remark enables us to obtain a discrete set of eigenvalues $\lambda_{lm}$.
In Eq. (16), $\frac{1}{2}M^{2}q^{2}\cos^{2}\theta$ can be treated
as a perturbation term on the generalized Legendre equation \cite{ABRAM}.
Thus, we obtain the following perturbation expansion:

\begin{equation}
\lambda_{lm}-\left(\frac{1}{2}\widehat{q}M\right)^{2}=\sum_{k=0}^{\infty}c_{k}\left(-1\right)^{k}\left(\frac{1}{2}\widehat{q}M\right)^{2k}.\label{17}
\end{equation}

It is worth noting that in Ref. \cite{ABRAM}, the expansion coefficients
in the summation symbol of Eq. (17), $\left\{ c_{k}\left(l,m\right)\right\} ,$
are explicitly given.

The radial equation of the KGF equation (14) in the MRLDBH geometry
acts as a Teukolsky equation \cite{TEU,TEU2}: 
\begin{equation}
\Delta^{2}\frac{d}{dr}\left(\Delta^{2}\frac{dR_{lm}}{dr}\right)+\left\{ \mathcal{H}^{2}-mM\left[2\widehat{q}Mr-mM+2h\left(r\right)\omega\right]-\Delta^{2}\left(K+h\left(r\right)\mu^{2}\right)\right\} R_{lm}=0,\label{18}
\end{equation}

where 
\begin{equation}
\Delta=r-M,\label{19}
\end{equation}

and 
\begin{equation}
\mathcal{H}=h\left(r\right)\omega+\frac{\widehat{q}}{2}\left(r^{2}+M^{2}\right).\label{20}
\end{equation}

For the MRLDBH, the event horizon is located at $r_{EH}=M,$ which
is the degenerate zero of Eq. (19).

In general (for asymptotically flat BHs), the bound nature of the
scalar clouds obey the following boundary conditions: purely ingoing
waves at the event horizon and decaying waves at spatial infinity
\cite{DDR,SDet,Dola,HodB1,HodB2}. However, when we study the limiting
behaviors of the radial equation (18), we get

\begin{equation}
R_{lm}\sim\begin{cases}
\frac{1}{\sqrt{r}}BesselJ(0,qe^{\frac{r^{\ast}}{r_{0}}})\approx\frac{1}{\sqrt{q}r}\sin\left(qe^{\frac{r^{\ast}}{r_{0}}}\right) & \text{ as }r\rightarrow\infty\ \ (r^{\ast}\rightarrow\infty),\\
\frac{1}{\sqrt{r_{EH}}}e^{-i\left[\omega-\left(m-\widehat{q}M\right)\widetilde{\Omega}\right]r^{\ast}} & \text{ as }r\rightarrow r_{EH}\ \ensuremath{r^{\ast}\rightarrow-\infty),\label{21}}
\end{cases}
\end{equation}

where $\widetilde{\Omega}=\left(r_{0}\right)^{-1}$ is the angular
velocity of $r_{EH},$ and $r^{\ast}$ is the tortoise coordinate
of the MRLDBH:

\begin{equation}
r^{\ast}=\int\frac{rr_{0}dr}{\Delta^{2}}=r_{0}\ln\left(\frac{\Delta}{\sqrt{2}}\right)-\frac{Mr_{0}}{\Delta}.\label{22}
\end{equation}

It is clear from the asymptotic solution (21) that unlike the Kerr BH \cite{HodSr},
there are no decaying (bounded) waves in MLRDBH geometry at spatial infinity. Instead
of this, we have oscillatory but fading waves (because of the factor
$\frac{1}{r}$) at the asymptotic region [see Fig. (1)]. This result probably originates
from the non-asymptotically flat structure of the MRLDBH geometry. 
\begin{figure}[h]
\centering
\includegraphics[scale=.65]{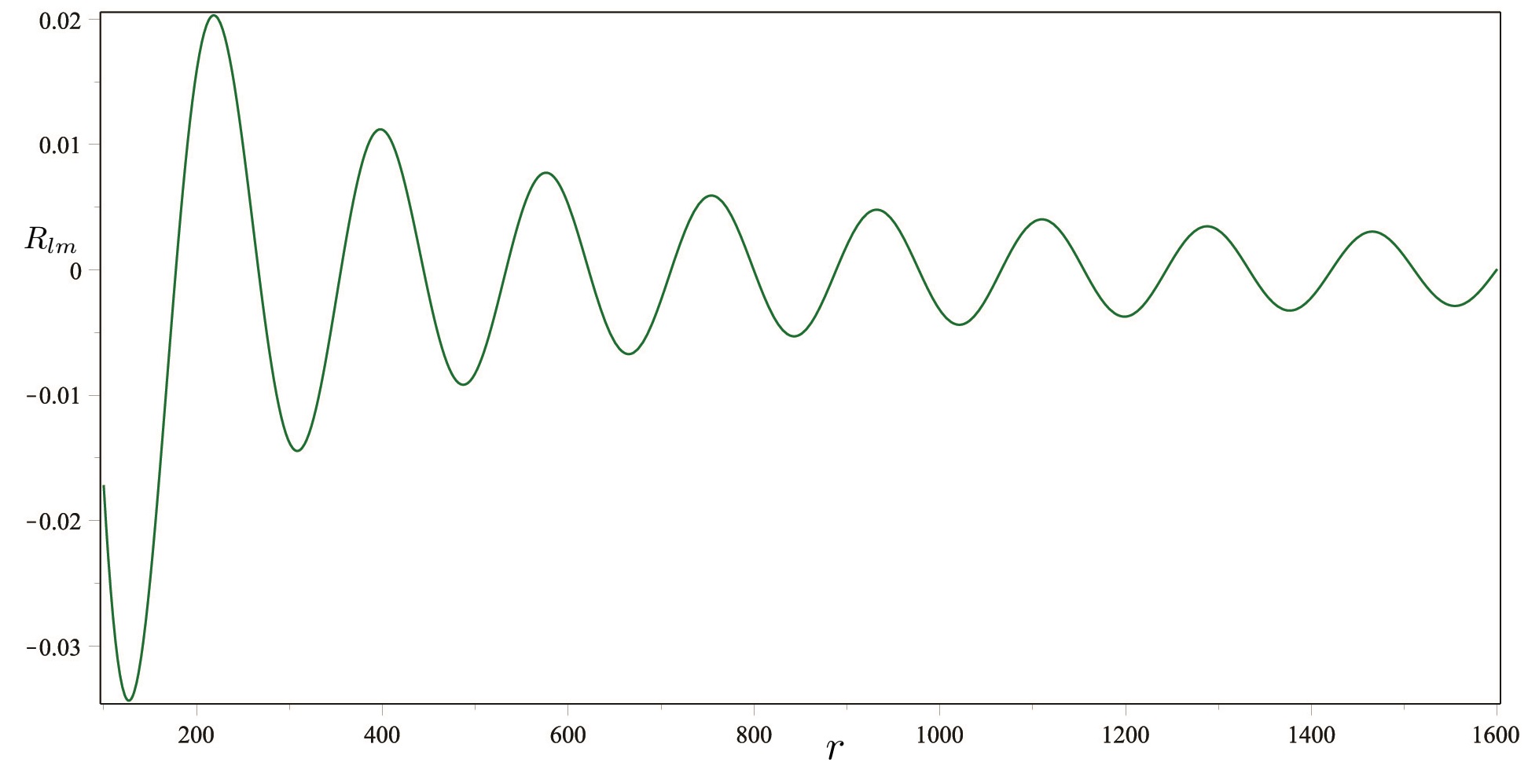}\caption{Asymptotic behavior of radial function $R_{lm}(r)$. The plot is governed by Eq. (21). The physical parameters are choosen as $M=1$, $q=0.05$, and $r_{0}=0.5$.}
\end{figure}

On the other hand, a very recent study \cite{confine} has shown that
scalar clouds can have a semipermeable surface, and thus they
might serve as a ``partial confinement'' in which only outgoing
waves are allowed to survive at the spatial infinity. Considering
this fact, we will impose a particular asymptotic boundary condition
to those non-decaying (unbounded) scalar clouds: only pure outgoing
waves propagate at spatial infinity.

\section{Stationary resonances and effective heights of the scalar clouds}

Stationary resonances or the so-called marginally stable modes\ are
the stationary regular solutions of the KGF equation (14) around the
horizon. They\ are characterized by $Im\left(\omega\right)=0$ \cite{HodSr},
which corresponds to $\omega=\left(m-\widehat{q}M\right)\widetilde{\Omega}$
for the MRLDBH spacetime. In fact, such resonances saturate the superradiant condition \cite{Clement}.

Introducing a new dimensionless variable

\begin{equation}
x=\frac{\Delta}{M},\label{23}
\end{equation}

one can see that the radial equation (18) can be rewritten as

\begin{equation}
x^{2}\frac{d^{2}R_{lm}}{dr^{2}}+2x\frac{dR_{lm}}{dr}+V_{eff}R_{lm}=0,\label{24}
\end{equation}

in which the effective potential is given by

\begin{equation}
V_{eff}=\left(\frac{1}{2}\widehat{q}Mx\right)^{2}+M\left(1+x\right)\left(m\widehat{q}-r_{0}{\mu}^{2}\right)+{m}^{2}-\lambda.\label{25}
\end{equation}

Letting

\begin{equation}
Y=xR_{lm}\text{ \ \ and \ \ }z=-i\widehat{q}Mx,\label{26}
\end{equation}

equation (24) transforms into the following differential equation

\begin{equation}
\frac{d^{2}Y}{dz^{2}}+\left(-\frac{1}{4}+\frac{\sigma}{z}+\frac{\frac{1}{4}-\beta^{2}}{z^{2}}\right)Y=0,\label{27}
\end{equation}

with

\begin{equation}
\sigma=i\left(m-\frac{r_{0}}{\widehat{q}}\mu^{2}\right)\text{ and }\beta^{2}=\lambda_{lm}+\frac{1}{4}-m^{2}+\left(r_{0}\mu^{2}-m\widehat{q}\right)M.\label{28}
\end{equation}

Without loss of generality, one can assume that $\beta$ is a non-negative
real number \cite{HodSr}. Therefore, Eq. (27) corresponds to a Whittaker
equation \cite{ABRAM}, whose solutions can be expressed in terms
of the confluent hypergeometric functions $\mathcal{M}(a,b,z)$ \cite{ABRAM,OLVER}.
Thus, the solution of Eq. (24) can be given by

\begin{equation}
R_{lm}=\frac{e^{-\frac{1}{2}z}}{\sqrt{z}}\left[C_{1}z^{\beta}\mathcal{M}\left(\frac{1}{2}+\beta-\sigma,1+2\beta,z\right)+C_{2}z^{-\beta}\mathcal{M}\left(\frac{1}{2}-\beta-\sigma,1-2\beta,z\right)\right],\label{29}
\end{equation}

where $C_{1}$ and $C_{2}$ are the integration constants. In the
vicinity of the horizon, solution (29) reduces to \cite{MOR}

\begin{equation}
R_{lm}\longrightarrow C_{1}z^{-\frac{1}{2}+\beta}+C_{2}z^{-\frac{1}{2}-\beta}.\label{30}
\end{equation}

Since the near horizon solution ($z\rightarrow0$) must admit the
regularity, one can figure out that

\begin{equation}
C_{2}=0\text{ and }\beta\geq\frac{1}{2}.\label{31}
\end{equation}

By using the asymptotic behaviors of the confluent hypergeometric
functions \cite{ABRAM,OLVER}, for $z\rightarrow\infty,$ Eq. (29)
can be approximated to

\begin{equation}
R_{lm}\longrightarrow C_{1}\left[e^{{\frac{1}{2}}z}{\frac{{\Gamma(1+2\beta)}}{{\Gamma({\frac{1}{2}}+\beta-\sigma)}}}z^{-1-\sigma}+e^{-{\frac{1}{2}}z}{\frac{{\Gamma(1+2\beta)}}{{\Gamma({\frac{1}{2}}+\beta+\sigma)}}}z^{-1+\sigma}(-1)^{-{\frac{1}{2}}-\beta+\sigma}\right].\label{32}
\end{equation}

Recalling the complex structure of $z$ {[}see Eq. (26){]}, we infer
that the first term in the square bracket of Eq. (32) stands for the
asymptotic ingoing waves, however the second one represents the asymptotic
outgoing waves. According to the physical boundary conditions aforementioned, the asymptotic ingoing wave $\left(\sim e^{{\frac{1}{2}}z}\right)$
in Eq. (32) must be terminated. This is possible by employing the
pole structure of the Gamma function \bigg($\Gamma(\tau)$ has the
poles at $\tau=-n$ for $n=0,1,2,...$\cite{ABRAM}\bigg). Therefore,
the resonance condition for the stationary unbound states of the field
eventually becomes

\begin{equation}
\frac{1}{2}+\beta-\sigma=-n.\label{33}
\end{equation}

It is convenient to express the radial solution of the unbound states
in a more compact form by using the generalized Laguerre polynomials
$L_{n}^{\left(2\beta\right)}\left(z\right)$ \cite{ABRAM}: 
\begin{equation}
R_{lm}=C_{1}z^{-\frac{1}{2}+\beta}e^{-{\frac{1}{2}}z}L_{n}^{\left(2\beta\right)}\left(z\right).\label{34}
\end{equation}

One can deduce from the resonance condition (33) that ${\sigma}$
should be a real number. However, taking cognizance of Eq. (28), which
indicates that ${\sigma}$ is a pure imaginary parameter, we conclude
that the resonances correspond to ${\sigma=0}$. Thus, we have two
cases:

\textit{Case-I}

\begin{equation}
m=\mu=0,\label{35}
\end{equation}

and \textit{Case-II}

\begin{equation}
\widehat{q}=\dfrac{r_{0}\mu^{2}}{m}.\label{36}
\end{equation}

It is worth noting that both cases exclude the existence of regular
\textit{static (}$\omega=0$\textit{)} solutions since $\omega=\left(m-\widehat{q}M\right)\widetilde{\Omega}$.
The latter remark is in accordance with the famous no-hair theorems
\cite{BekT2,sp13,sp14,sph1,sph2,NH1,NH4,NH6,NH7,NH8} since they exempt
the static hairy configurations \cite{sp01}.

For solving the resonance condition (33), it is practical to introduce
another dimensionless variable:

\begin{equation}
\epsilon=\frac{i}{2}\widehat{q}M,\label{37}
\end{equation}

so that Eq. (28) can be rewritten as \cite{HodSr,ABRAM}

\begin{equation}
{\sigma}=\frac{\sqrt{2}M\mu^{2}Q+2im\epsilon}{2\epsilon},\label{38}
\end{equation}

\begin{equation}
\beta^{2}=\left(l+\frac{1}{2}\right)^{2}-m^{2}-\epsilon^{2}+M\mu^{2}r_{0}+2im\epsilon+\sum\limits _{k=1}^{\infty}c_{k}\epsilon^{2k}.\label{39}
\end{equation}

After substituting Eqs. (38) and (39) into Eq. (33), we express the
resonance condition as a polynomial equation for $\epsilon$:

\[
8\left[l\left(l+1\right)+m^{2}-1\right]\epsilon^{4}+\left(2l-1\right)\left(2l+3\right)\left\{ -8im\epsilon^{3}-4\left[im(2n+1)-n\left(n+1\right)+l\left(l+1\right)+r_{0}M\mu^{2}\right]\epsilon^{2}\right.
\]

\begin{equation}
\left.+2r_{0}M\mu^{2}\left[2im-\left(2n+1\right)\right]\epsilon+\left(r_{0}M\mu^{2}\right)^{2}-4\sum\limits _{k=2}^{\infty}c_{k}\epsilon^{2k+2}\right\} =0.\label{40}
\end{equation}

Discrete and infinite group of stationary resonances for both cases
are presented in Tables I and II. The results are shown for different
values of $n$ (resonance parameter). Unlike the Kerr BH \cite{HodSr}, our numerical calculations about Eq. (40) showed that in the case of $n=0$ the obtained resonance
values $(Mq_{\text{resonance}})$ are complex,
which do not admit physically acceptable results. For this reason,
we consider the resonance parameters of having $n\geq1$.

\begin{table}[htbp]
\centering %
\begin{tabular}{|c|c|c|c|}
\hline 
$n$  & $Mq_{\text{resonance}}$  & $\left\vert x\right\vert _{\text{cloud}}^{\left(n\right)}$  & \tabularnewline
\hline 
\ 1\ \  & \ 2.4495\ \  & \ 1.7321  & \tabularnewline
\ 2\ \  & \ 4.2426\ \  & \ 1.6667  & \tabularnewline
\ 3\ \  & \ 6.0000\ \  & \ 1.6499  & \tabularnewline
\ 4\ \  & \ 7.7460\ \  & \ 1.6432  & \tabularnewline
\hline 
\end{tabular}\caption{Stationary scalar resonances of a MRLDBH for \textit{Case-I}. The
values of $Mq_{\text{resonance}}$ and the effective heights of the hairosphere for a 
\textit{s-wave} ($l=m=0$) with increasing resonance parameter $n\geq1$
are tabulated.}
\label{Table1} 
\end{table}

\bigskip{}

\begin{table}[htbp]
\centering %
\begin{tabular}{|c|c|c|c|}
\hline 
$n$  & $Mq_{\text{resonance}}$  & $\left\vert x\right\vert _{\text{cloud}}^{\left(n\right)}$  & \tabularnewline
\hline 
\ 1\ \  & \ 1.5811\ \  & \ 2.6833  & \tabularnewline
\ 2\ \  & \ 3.5355\ \  & \ 2.0000  & \tabularnewline
\ 3\ \  & \ 5.2440\ \  & \ 1.8878  & \tabularnewline
\ 4\ \  & \ 6.8920\ \  & \ 1.8468  & \tabularnewline
\hline 
\end{tabular}\caption{Stationary scalar resonances of a MRLDBH for \textit{Case-II}. The
values of $Mq_{\text{resonance}}$ and the effective heights of the
hairosphere are represented for the fundamental resonances $l=m=1$
with increasing resonance parameter $n\geq1$. Meanwhile, although
it is not depicted here, qualitatively similar behaviors are observed
for the other values of $\{l,m\}$.}
\label{Table2} 
\end{table}

We now consider the effective heights of the stationary charged massive
scalar field configurations surrounding the MRLDBH. These configurations
correspond to the group of wave-functions (34) that fulfill the resonance
condition (33). According to the `\textit{no short hair theorem}'
proposed for the spherically symmetric and static hairy BH configurations
\cite{NH6}, the hairosphere \cite{hairo} must extend beyond $\frac{3}{2}r_{EH}$. Taking cognizance
of Eq. (23), we conclude that the \textit{minimum} radius of the hairosphere
corresponds to $\left\vert x\right\vert _{\text{hair}}=\frac{1}{2},$
where $\left\vert x\right\vert $ is the dimensionless height (absolute
altitude). Furthermore, we can compute the size of the stationary
scalar clouds by defining their effective radii. The effective heights
of the scalar clouds can be approximated to a radial position at which
the quantity $4\pi\left\vert x\right\vert ^{2}\left\vert \Psi\right\vert ^{2}$
reaches its global maximum value \cite{HodSr,communication}. By using
Eq. (34), one finds the dimensionless heights of the clouds, as follows:

\begin{equation}
\left\vert x\right\vert _{\text{cloud}}^{\left(n\right)}=\left\vert \frac{2\beta+1+2n}{2\epsilon}\right\vert ,\text{ \ \ \ \ }n=1,2,3,......
\end{equation}
The effective heights of the principal clouds above the central BH
for both Case I and Case II are displayed in Table I and Table II,
respectively. It can be easily seen that \{$\left\vert x\right\vert _{\text{cloud}}^{\left(n\right)}$\}
are always larger than the lower bound ($\left\vert x\right\vert _{\text{hair}}\geq\frac{1}{2}$)
of the hairosphere \cite{HodSr}. 

\section{Conclusion}

In this study, we have explored the dynamics of a charged massive
scalar field in the background of MRLDBH. It has been shown that there
exists a quantized and infinite set of resonances that describes non-decaying
charged massive scalar configurations (clouds) enclosing the MRLDBH.
We have analytically computed the effective heights of the hairosphere and shown that $\left\vert x\right\vert _{\text{hair}}\geq\frac{1}{2}$. At this juncture, one can interrogate our findings about whether they are compatible with  \textit{no-short hair} theorem \cite{hairo} or not. Because, in the seminal works of Hod \cite{nsht1,nsht2}, it was discussed that \textit{charged} rotating black holes
can have short bristles, and thus they provide evidence for the failure
of no-short hair theorem. However, his another and most recent work \cite{Hodcqg} has supported the no-short hair theorem: external
matter fields of a static spherically symmetric rotating hairy black
hole (\textit{Kerr BH case}) configuration must extend beyond the
null circular geodesic which characterizes the corresponding BH
spacetime. In fact, rotating linear dilaton BHs show remarkable similarities to the Kerr BH, instead of its charged version: Kerr-Newman BH. This is because of their \textit{fixed} background charge $Q=\frac{r_{0}}{\sqrt{2}}$, which is not existed in the horizons [see Eqs. (5) and (6)]. Furthermore, it tunes the radius of the spherical part of the metric (2). Namely, unlike the Kerr-Newman BH ( $Q\rightarrow0$ reduces it to the Kerr BH), a rotating linear dilaton BH
has no zero-charge limit [see Eqs. (2-4)]. This point was highlighted in the original paper of the rotating linear dilaton BHs \cite{Clement}. So, similar to \cite{HodSr,Hodcqg}, our results give also support to the no-short hair theorem.

In conclusion, our analytical findings for MRLDBHs support the existence
of non-decaying scalar field dark matter halo around the rotating
BHs. In particular, we have shown that the unbound-state resonances are distinctively
possible with two cases: \textit{Case-I} (35) and \textit{Case-II}
(36). In both cases, $Mq_{\text{resonance}}$ spectrum of the MRLDBH
does not hold for the ground state resonances with $n=0$. Therefore,
we have considered the resonances for $n\geq1$. The analytically
derived values of $Mq_{\text{resonance}}$ are numerically illustrated
in Tables I and II for the fundamental resonances $l=m=0$ and $l=m=1$,
respectively. It is worth noting that the other combinations of $\{l,m\}$
values give almost the same results. 

It would be interesting to extend this study for the dynamics of a
charged massive field having spins other than zero. We will focus
on this in our next research in the near future. 

\section*{Acknowledgements}

We would like to thank Prof. Shahar Hod, Prof. Carlos A.R. Herdeiro, and Prof. Canisius Bernard for suggestions
and correspondences.


\begin{thebibliography}{10}
\bibitem{communication} According to the private communications between
Hod (the author of \cite{HodB2}) and us, it is understood that there
is a typo in the expression of \char`\"{}\textit{effective radii as
the radii at which the quantity }$4\pi r^{2}\left\vert \Psi\right\vert ^{2}$\textit{\ attains
its global maximum}\char`\"{} of \cite{HodB2}: In the expression
of $4\pi r^{2}\left\vert \Psi\right\vert ^{2}$, $r$ should be replaced
with the dimensionless height $\left\vert x\right\vert $.
\end{thebibliography}
\end{document}